\documentclass[12pt]{article}
\usepackage{amsfonts}
\usepackage{latexsym}
\usepackage{amsmath}
\usepackage{amssymb}
\usepackage{amssymb}

\newcommand{\newsection}{    
\setcounter{equation}{0}\section}
\def\appendix#1{\addtocounter{section}{1}\setcounter{equation}{0}
\renewcommand{\thesection}{\Alph{section}}
\section*{Appendix \thesection\protect\indent \parbox[t]{11.15cm}{#1}}
\addcontentsline{toc}{section}{Appendix \thesection\ \ \ #1}}

\newcommand{\be}{\begin{eqnarray}}
\newcommand{\ee}{\end{eqnarray}}
\newcommand{\bea}{\begin{eqnarray}}
\newcommand{\eea}{\end{eqnarray}}
\newcommand{\ba}{\begin{array}}
\newcommand{\ea}{\end{array}}
\newcommand{\nn}{\nonumber \\}

\newenvironment{frcseries}{\fontfamily{frc}\selectfont}{}

\def \la {\label}

\def\a{\alpha}
\def\b{\beta}
\def\l{\lambda}
\def\g{\gamma}

\def\e{\epsilon}

\def\hn{{\hat{\nu}}}

\def\bbe{{\bf{e}}}
\font\mybb=msbm10 at 11pt

\def\bb#1{\hbox{\mybb#1}}

\def\bZ {\bb{Z}}
\def\bR {\bb{R}}

\def\bH {\bb{H}}
\def\bC {\bb{C}}

\def\tn {{\tilde{\nabla}}}

\def\hn {{\hat{\nabla}}}

\begin{document}
\begin{titlepage}
\begin{center}

\vspace*{-1.0cm}
\hfill DMUS--MP--13/09 \\

\vspace{2.0cm} {\Large \bf  IIB horizons} \\[.2cm]

\vspace{1.5cm}
 {\large  U. Gran$^1$, J. Gutowski$^2$ and  G. Papadopoulos$^3$}

\vspace{0.5cm}

${}^1$ Fundamental Physics\\
Chalmers University of Technology\\
SE-412 96 G\"oteborg, Sweden\\

\vspace{0.5cm}
$^2$ Department of Mathematics \\
University of Surrey \\
Guildford, GU2 7XH, UK \\

\vspace{0.5cm}
${}^3$ Department of Mathematics\\
King's College London\\
Strand\\
London WC2R 2LS, UK\\

\vspace{0.5cm}

\end{center}

\vskip 1.5 cm
\begin{abstract}
 We solve the Killing spinor equations for all near-horizon IIB geometries which preserve at least one supersymmetry. We
 show that  generic horizon sections are 8-dimensional almost Hermitian spin${}_c$ manifolds. Special cases include horizon sections with a
 $Spin(7)$ structure and those for which the Killing spinor is pure.
We also explain how the common sector horizons and the horizons with only 5-form flux are included in our general analysis.
  We investigate several special cases
mainly focusing on the horizons with constant scalars admitting a pure Killing spinor and find that some of these exhibit a generalization of the 2-SCYT condition that arises
 in the horizons with 5-form fluxes only. We use this to construct new examples of near-horizon geometries with both 3-form and 5-form fluxes.

\end{abstract}

\end{titlepage}



\setcounter{section}{0}
\setcounter{subsection}{0}
\setcounter{equation}{0}

\newsection{Introduction}

In understanding the physics of black holes and branes, and in AdS/CFT,  horizons have played a central role. In most applications, the horizons are Killing\footnote{This means that they admit a time Killing vector field which becomes null at a spacetime  hyper-surface.}, and in the extreme case can be investigated in a unified way because there is a local model for the near-horizon  fields \cite{isen, gnull}. Typically,  some additional assumptions are made depending on the problem.  The most common assumption is that the metric and all the other form field strengths\footnote{Non-smooth near-horizon geometries can also be investigated
in the special case for which there exists a frame
 in which the metric is smooth. This class includes all branes.  In this case, one works in a  frame with
 respect to which the metric is smooth, and allow singularities in the
other fields, such as the scalars of the theory. If all fields are smooth, the choice of frame is not essential.}  are taken to be smooth.
This assumption includes many black hole horizons, as well as
the horizons of some branes, like the D3-, M2- and M5-branes.  Therefore it is expected that a systematic understanding of all near-horizon geometries will have applications
in AdS/CFT \cite{maldacena} as well as in the construction of IIB back hole solutions which exhibit horizons with exotic topology and geometry, see e.g.~\cite{israel}-\cite{ring} for earlier related works.

In this paper, we solve the Killing spinor equations (KSEs) of IIB supergravity for all near-horizon geometries preserving at least one supersymmetry and determine the geometry of the horizon sections.
This extends the results on M-horizons \cite{mhor} and heterotic horizons \cite{hhor} to IIB supergravity. In the investigation of the KSEs for IIB horizons three cases\footnote{This is similar to the three cases that arise in the solution of the KSEs for IIB backgrounds preserving one supersymmetry in \cite{sgiib}.} arise depending on the choice of Killing spinor
 (i) the generic horizons, (ii) the $Spin(7)$ horizons and (iii) the pure $SU(4)$ horizons. We find that the sections ${\cal S}$ of generic
near-horizon geometries are 8-dimensional almost Hermitian spin${}_c$ manifolds, and the sections of the $Spin(7)$ horizons admit  a $Spin(7)$ structure.
The sections of the pure $SU(4)$ horizons, which admit  a pure Killing spinor, are again 8-dimensional almost Hermitian spin${}_c$ manifolds. In all the above cases,
we also examine the implications on the topology and geometry of horizon sections depending on whether the IIB scalars take values in the hyperbolic upper half-plane
or the hyperbolic upper half-plane after an $SL(2,\bZ)$ U-duality identification.

We further investigate near-horizon geometries with constant scalars.  The structure group of horizon sections both for generic  and pure $SU(4)$ horizons
reduces further to a subgroup of $SU(4)$.  Therefore in these two cases, the near-horizon sections are almost Hermitian spin${}_c$ manifolds with an $SU(4)$ structure.
 In all IIB horizons  without further restrictions on the fields, the KSEs do not impose  additional restrictions\footnote{Of course to find solutions, one has to impose in addition the Bianchi identities and field equations.} on the geometry of the horizon sections.
We also explore several examples mostly focusing on the pure $SU(4)$ horizons with constant scalars. In the complex case,  we find a deformation of the 2-SKT
condition, $\partial\bar\partial\omega^2=0$, of horizons with only 5-form fluxes \cite{iibhorf5} with the addition of  a source term which depends on the 3-form fluxes. We use this to give new
examples of horizons with 3- and 5-form fluxes.  We also describe how the horizons with only 5-form flux and the horizons of the common sector arise as special
cases of our IIB horizons.

To prove the above results, we first demonstrate that the KSEs are integrable along the
lightcone directions giving rise to a system of differential and algebraic equations on the horizon section ${\cal S}$. We find using the field equations, Bianchi identities and the bilinear matching condition\footnote{The bilinear matching condition is the identification
of the stationary Killing vector field of a black hole with the Killing spinor vector bilinear.} that  this
 system can be considerably  simplified giving rise to a parallel transport equation and an algebraic equation on the horizon sections
 associated with the gravitino and dilatino KSEs, respectively. These two equations are solved using spinorial geometry \cite{spgeom}, and as we have already mentioned, there are three cases to consider:
 the generic horizons, the $Spin(7)$ horizons, and the horizons with a pure Killing spinor.  The geometry of the horizon sections is next investigated by computing the Killing spinor
 bilinears, and in particular those which are associated with nowhere vanishing forms on the horizon sections.
 In the analysis above, the compactness properties of section ${\cal S}$ have not been used. Because
 of this, the results apply to  both brane and black hole horizons. However in the construction of examples, like those associated with the deformation of the 2-SKT condition,
  we take ${\cal S}$  to be compact without boundary as expected
 for black hole horizon sections.

 This paper has been organized as follows. In section 2, we state the near-horizon metric and the form fluxes of IIB supergravity and
 decompose the field equations and Bianchi identities along the lightcone and horizon section ${\cal S}$ directions. Moreover we derive
 which equations are independent.  In section 3, we integrate the KSEs along the lightcone directions and use the field equations, Bianchi identities
 and the bilinear matching condition to derive the independent KSEs. In section 4, we solve the independent KSEs using spinorial geometry and
 express the fluxes in terms of the geometry.  Moreover, we derive the geometry of the horizon section and describe some aspects of their topology. In section 5,
 we explain how the horizons with only 5-form flux and those of the common sector are included as special cases in our analysis. In section 6, we explore the
 geometry of horizon sections admitting a  pure Killing spinor and demonstrate that in the complex case the 2-SKT conditions of 5-form horizons is deformed
 with the 3-form fluxes. In section 7, we state our conclusions, and in appendix A we give the solution of the linear system associated with IIB horizons.

\newsection{Fields and dynamics near a horizon}

\subsection{Fields and KSEs}

To describe the form field strengths of IIB supergravity, consider the complex line bundle $\lambda$ over the spacetime. This  is the pull-back of the canonical
bundle of either the coset space $\bH=SU(1,1)/U(1)$ or after a U-duality identification of  $SL(2,\bR)\backslash \bH$ with respect to the dilaton and axion, the two scalar fields of the theory.  $\bH$ is identified with the hyperbolic upper half-plane which is a manifold and   $SL(2,\bR)\backslash \bH$ is identified with the fundamental domain of the moduli space of complex structures of a 2-torus which is an orbifold.

The field strengths of the bosonic fields of IIB supergravity are  a $\lambda^2$-valued complex 1-form $P$, a $\lambda$-valued complex 3-form $G$ and a
self-dual 5-form $F$. $P$, $G$ and $F$ are the field strengths of two scalars, two 2-form gauge potentials and a 4-form gauge potential, respectively.
Assuming that the horizons under investigation are killing horizons,  and under some regularity assumptions, the IIB
fields in the extreme near-horizon limit can be expressed as
\be
ds^2 =2 \bbe^+ \bbe^- + \delta_{ij} \bbe^i \bbe^j = 2 du (dr + rh -{1 \over 2} r^2 \Delta du)+ \gamma_{IJ} dy^I dy^J~,
\la{hormetr}
\ee
\be
\label{fivef}
F= r \bbe^+ \wedge X + \bbe^+ \wedge \bbe^- \wedge Y + \star_8 Y~,
\ee
\bea
G = r \bbe^+ \wedge L + \bbe^+ \wedge \bbe^- \wedge \Phi + H~,
\label{threef}
\eea
\bea
P = \xi~,
\label{onef}
\eea
where we have introduced the frame
\be
\label{basis1}
\bbe^+ = du, \qquad \bbe^- = dr + rh -{1 \over 2} r^2 \Delta du, \qquad \bbe^i = e^i_I dy^I~,
\ee
and used the self-duality\footnote{We choose $\epsilon_{0123456789}=1$ and the spacetime volume form is related to that of the horizon section as $\mathrm{dvol}(M)
=\bbe^+\wedge \bbe^-\wedge \mathrm{dvol}({\cal S})$. In particular, $(\star_8 Y)_{n_1 n_2 n_3 n_4 n_5} = {1 \over 3!} \epsilon_{n_1 n_2 n_3 n_4 n_5}{}^{m_1 m_2 m_3} Y_{m_1 m_2 m_3}$.} of $F$,
\be
F_{M_1 M_2 M_3 M_4 M_5} =-{1 \over 5!} \epsilon_{M_1 M_2 M_3 M_4 M_5}{}^{N_1 N_2 N_3 N_4 N_5}
F_{N_1 N_2 N_3 N_4 N_5}~,
\ee
which also requires that
\be
\label{sd1}
X = - \star_8 X~.
\ee
IIB supergravity also admits a $U(1)$ connection of $\lambda$, denoted by $Q$. Upon restriction to the near-horizon geometry,
\bea
Q = \Lambda ~,
\eea
where $\Lambda$ is $r,u$-independent, i.e. $Q_+=Q_-=0$,  and so $\Lambda$ is a connection of $\lambda$ restricted to ${\cal S}$.

The dependence on the coordinates $u$ and $r$ is explicitly given, and so $h$, $\Delta$, $X$, $Y$, $L$, $\Phi$, $H$, $\Lambda$ and $\xi$ depend only
on the coordinates $y^I$ of the horizon section ${\cal S}$ which is the co-dimension 2 subspace given by  $r=u=0$. In particular, the scalars depend only on $y$, and  $\xi$ is a section of $\lambda^2\otimes \Lambda^1({\cal S})$, where we have restricted\footnote{Note that $\lambda$ is topologically trivial if it is the pull-back of a line bundle over the hyperbolic upper half-plane which is a contractible space.   This will also be the case after an $SL(2, \bZ)$ U-duality identification of the IIB scalars as the j-function maps the fundamental domain  homeomorphically  to $\bC$,  unless
two copies are glued together to a sphere as in the case of 24 cosmic strings \cite{vafa}. In both cases, $\lambda$ may be equipped with a connection with non-vanishing curvature.}  $\lambda$ to ${\cal S}$. Similarly, $L$, $\Phi$ and $H$ are sections of $\lambda\otimes \Lambda^2({\cal S})$,  $\lambda\otimes \Lambda^1({\cal S})$ and  $\lambda\otimes \Lambda^3({\cal S})$, respectively, and $X$ and $Y$ are sections of $\Lambda^4({\cal S})$ and  $\Lambda^3({\cal S})$, respectively.
Observe that $V={\partial \over \partial u}$ is the horizon  timelike Killing vector which becomes null at $r=0$.  The derivation of (\ref{hormetr})-(\ref{onef}) is similar to that of other
near-horizon geometries with form fluxes and we shall not elaborate  here.

\subsection{Bianchi identities and field equations}

Before proceeding to an analysis of the KSEs, it is useful to evaluate the Bianchi identities and the field equations of IIB supergravity \cite{schwarz1, schwarz2}
on the near-horizon fields (\ref{hormetr})-(\ref{onef}).
First consider the Bianchi identity
\bea
\label{bian1}
dF -{i \over 8} G \wedge {\bar{G}}=0~.
\eea
This implies that
\bea
\label{ba1}
X = d_hY -{i \over 8} (\Phi \wedge {\bar{H}} - {\bar{\Phi}} \wedge H)~,~~~d_hY=dY-h\wedge Y~,
\eea
\bea
\label{ba2}
d \star_8 Y = {i \over 8} H \wedge {\bar{H}}~,
\eea
or equivalently
\bea
\label{ba2b}
\tn^\ell Y_{\ell ij} = -{i \over 288} \epsilon_{ij}{}^{\ell_1 \ell_2 \ell_3 \ell_4 \ell_5 \ell_6} H_{\ell_1 \ell_2 \ell_3}
{\bar{H}}_{\ell_4 \ell_5 \ell_6}~,
\eea
where $\tn$ is the Levi-Civita connection on ${\cal S}$.
Therefore $X$ is not independent. However, notice  that the anti-self duality condition  (\ref{sd1}) of $X$ imposes an additional restriction
on the components of the fluxes that appear in (\ref{ba1}).

Next, the Bianchi identity
\bea
\label{bian2}
dG -i Q \wedge G + P \wedge {\bar{G}}=0~,
\eea
implies that
\bea
\label{ba3}
L = d_h \Phi  -i \Lambda \wedge \Phi + \xi \wedge {\bar{\Phi}}~,
\eea
and
\bea
\label{ba4}
dH = i \Lambda \wedge H - \xi \wedge {\bar{H}}~.
\eea
So again $L$ is not independent and can be expressed in terms of the other fluxes.

The Bianchi identity
\bea
\label{bian3}
dP -2i Q \wedge P =0~,
\eea
implies
\bea
\label{ba5}
d \xi = 2i \Lambda \wedge \xi~,
\eea
and the Bianchi identity
\bea
\label{bian4}
dQ =-i P \wedge {\bar{P}}~,
\eea
implies
\bea
\label{ba6}
d \Lambda = -i \xi \wedge {\bar{\xi}}~.
\eea

We remark that ({\ref{bian1}}) and ({\ref{bian2}}) imply two additional identities involving $dX$ and $dL$,
however, these are in fact implied by ({\ref{ba1}}), ({\ref{ba2}}), ({\ref{ba3}}), ({\ref{ba4}}), ({\ref{ba5}}), ({\ref{ba6}}). As a consequence, the latter
are  necessary and sufficient conditions imposed by the Bianchi identities.

Next we turn to the bosonic field equations.  Substituting the near-horizon fields into the 2-form gauge potentials field equations
\bea
\label{ggauge1}
\nabla^C G_{ABC} -i Q^C G_{ABC} - P^C {\bar{G}}_{ABC} +{2i \over 3} F_{AB N_1 N_2 N_3} G^{N_1 N_2 N_3} =0~,
\eea
one finds that
\bea
\label{feq1}
\tn^i \Phi_i -i \Lambda^i \Phi_i - \xi^i {\bar{\Phi}}_i +{2 i \over 3} Y_{\ell_1 \ell_2 \ell_3} H^{\ell_1 \ell_2 \ell_3}=0~,
\eea
\bea
\label{feq2}
- \tn^i L_{im} +i \Lambda^i L_{im} + h^i L_{im} -{1 \over 2} dh^{ij} H_{ijm}
+ \xi^i {\bar{L}}_{im} +{2i \over 3} (X_{m \ell_1 \ell_2 \ell_3} H^{\ell_1 \ell_3 \ell_3}
-3 Y_{m \ell_1 \ell_2} L^{\ell_1 \ell_2} ) =0~,
\nn
\eea
and
\bea
\label{feq3}
\tn^\ell H_{\ell ij} -i \Lambda^\ell H_{\ell ij}- h^\ell H_{\ell ij} + L_{ij} - \xi^\ell {\bar{H}}_{\ell ij}
+{2i \over 3}( \star_8 Y_{ij \ell_1 \ell_2 \ell_3} H^{\ell_1 \ell_2 \ell_3} - 6 Y_{ij \ell} \Phi^\ell) =0~.
\eea
Similarly, the field equation of the scalars
\bea
\label{scalar1}
\nabla^A P_A -2i Q^A P_A +{1 \over 24} G_{N_1 N_2 N_3} G^{N_1 N_2 N_3} =0~,
\eea
implies
\bea
\label{feq4}
\tn^i \xi_i -2i \Lambda^i \xi_i - h^i \xi_i +{1 \over 24}(-6 \Phi^i \Phi_i + H_{\ell_1 \ell_2 \ell_3} H^{\ell_1 \ell_2 \ell_3}) =0~.
\eea
Note that there is no independent field equation for $F$ because $F$ is self-dual.

It remains to investigate the Einstein equation
\bea
\label{ein1}
R_{AB} - {1 \over 6} F_{A L_1 L_2 L_3 L_4} F_B{}^{L_1 L_2 L_3 L_4} -{1 \over 4} G_{(A}{}^{N_1 N_2} {\bar{G}}_{B) N_1 N_2}
\nn
+{1 \over 48} g_{AB} G_{N_1 N_2 N_3} {\bar{G}}^{N_1 N_2 N_3} -2 P_{(A} {\bar{P}}_{B)}=0~.
\eea
Substituting the near-horizon fields, one finds that  $+-$ component gives
\bea
\label{feq5}
{1 \over 2} \tn^i h_i - \Delta - {1 \over 2} h^2 + {2 \over 3} Y_{\ell_1 \ell_2 \ell_3} Y^{\ell_1 \ell_2 \ell_3}
+{3 \over 8} \Phi^i {\bar{\Phi}}_i +{1 \over 48} H_{\ell_1 \ell_2 \ell_3} {\bar{H}}^{\ell_1 \ell_2 \ell_3} =0~.
\eea
Similarly, the $+i$, $ij$ and $++$ components imply that
\bea
\label{feq6}
-{1 \over 2} \tn^j dh_{ji} - dh_{ij} h^j - \tn_i \Delta + \Delta h_i
+{4 \over 3} X_{i \ell_1 \ell_2 \ell_3} Y^{\ell_1 \ell_2 \ell_3}
\nn
-{1 \over 8} \bigg( L_{\ell_1 \ell_2} {\bar{H}}_i{}^{\ell_1 \ell_2}
-2 \Phi^\ell {\bar{L}}_{i \ell} + {\bar{L}}_{\ell_1 \ell_2} H_i{}^{\ell_1 \ell_2}
-2 {\bar{\Phi}}^\ell L_{i \ell} \bigg) =0~,
\eea
\bea
\label{feq7}
{\tilde{R}}_{ij} + \tn_{(i} h_{j)} -{1 \over 2} h_i h_j +4 Y_{i \ell_1 \ell_2} Y_j{}^{\ell_1 \ell_2}
+{1 \over 2} \Phi_{(i} {\bar{\Phi}}_{j)} -2 \xi_{(i} {\bar{\xi}}_{j)}
-{1 \over 4} H_{\ell_1 \ell_2 (i} {\bar{H}}_{j)}{}^{\ell_1 \ell_2}
\nonumber \\
+ \delta_{ij} \bigg(-{1 \over 8} \Phi_\ell {\bar{\Phi}}^\ell -{2 \over 3} Y_{\ell_1 \ell_2 \ell_3} Y^{\ell_1 \ell_2 \ell_3}
+{1 \over 48} H_{\ell_1 \ell_2 \ell_3} {\bar{H}}^{\ell_1 \ell_2 \ell_3} \bigg) =0~,
\eea
and
\bea
\label{feq8}
{1 \over 2} \tn^2 \Delta -{3 \over 2} h^i \tn_i \Delta -{1 \over 2} \Delta \hn^i h_i + \Delta h^2
+{1 \over 4} dh_{ij} dh^{ij} -{1 \over 6} X_{\ell_1 \ell_2 \ell_3 \ell_4} X^{\ell_1 \ell_2 \ell_3 \ell_4}
-{1 \over 4} L_{ij} {\bar{L}}^{ij} =0~,
\nonumber \\
\eea
respectively, where $\tilde R$ is the Ricci tensor of ${\cal S}$.

Not all of the above field equations are independent. In particular,
on taking the divergence of ({\ref{feq3}}), and making use of ({\ref{ba1}}), ({\ref{ba2}}), ({\ref{ba3}}), ({\ref{ba4}}),
({\ref{ba5}}), ({\ref{ba6}}), ({\ref{feq3}}), ({\ref{sd1}}), one obtains ({\ref{feq2}}).
Also, on computing the Einstein tensor of ${\cal{S}}$ using ({\ref{feq7}}) and ({\ref{feq5}}),
and evaluating the Einstein tensor Bianchi identity, one obtains ({\ref{feq6}}) upon making use
of  ({\ref{ba1}}), ({\ref{ba2}}), ({\ref{ba3}}), ({\ref{ba4}}), ({\ref{ba5}}), ({\ref{feq1}}), ({\ref{feq3}}),
({\ref{feq4}}).
Furthermore, on taking the divergence of ({\ref{feq6}}), one also obtains ({\ref{feq8}}), after
making use of ({\ref{sd1}}) and ({\ref{ba1}}) to compute $d \star_8 X$, ({\ref{ba3}}) to compute $dL$,
and also ({\ref{feq6}}) and ({\ref{feq3}}).
Hence it follows that ({\ref{feq2}}), ({\ref{feq6}}) and ({\ref{feq8}}) are implied by the other field equations and Bianchi identities.

\setcounter{equation}{0}
\section{KSEs on IIB horizons}

\subsection{Lightcone integrability of KSEs}

The gravitino and dilatino KSEs of IIB supergravity \cite{schwarz1, schwarz2} are
\bea
\label{gkse}
&&\bigg(\nabla_M -{i \over 2} Q_M
+{i \over 48} F_{M N_1 N_2 N_3 N_4}\Gamma^{N_1 N_2 N_3 N_4} \bigg) \epsilon
\cr
&&~~~~~~~~~~~~-{1 \over 96} \bigg(\Gamma_M{}^{N_1 N_2 N_3} G_{N_1 N_2 N_3} -9 G_{M N_1 N_2} \Gamma^{N_1 N_2} \bigg) C* \epsilon
=0~,
\eea
\bea
\label{akse}
P_M \Gamma^M C* \epsilon +{1 \over 24} G_{N_1 N_2 N_3} \Gamma^{N_1 N_2 N_3} \epsilon =0~.
\eea
For our spinor conventions see \cite{sgiib}.

As in the analysis of KSEs for near-horizon geometries in other supergravity theories, the IIB KSEs can be integrated along the lightcone directions.
In particular solving the $-$ component of the gravitino KSE, one finds
\bea
\label{st1}
\epsilon_+ &=& \phi_+
\nn
\epsilon_-&=&\phi_- + r \Gamma_- \bigg({1 \over 4} h_i \Gamma^i
+{i \over 12} Y_{n_1 n_2 n_3} \Gamma^{n_1 n_2 n_3} \bigg) \phi_+
\nn
 &+&r \Gamma_-\bigg( {1 \over 96} H_{\ell_1 \ell_2 \ell_3} \Gamma^{\ell_1 \ell_2 \ell_3} +{3 \over 16} \Phi_i \Gamma^i \bigg) C* \phi_+~,
\eea
where $\phi_\pm$ do not depend on $r$. Similarly, the solution of the  $+$ component of the gravitino KSE gives
\bea
\label{st2}
\phi_+ &=&\eta_+ + u \Gamma_+ \bigg({1 \over 4} h_i \Gamma^i
-{i \over 12} Y_{n_1 n_2 n_3} \Gamma^{n_1 n_2 n_3} \bigg) \eta_- ~,
\nn
&+& u \Gamma_+ \bigg({1 \over 96} H_{\ell_1 \ell_2 \ell_3} \Gamma^{\ell_1 \ell_2 \ell_3}
-{3 \over 16} \Phi_i \Gamma^i \bigg) C* \eta_-
\nn
\phi_- &=& \eta_- ~,
\eea
where $\eta_\pm$ do not depend on the $u$ and $r$ coordinates. Furthermore, $\eta_+$ and $\eta_-$ must also satisfy
a number of algebraic conditions. To simplify these, we shall first identify the 1-form associated with the
 stationary Killing vector field $V=\partial_u$ of horizons with the 1-form $Z$ constructed as a Killing spinor bilinear.

\subsubsection{Vector bilinear matching condition} \la{vbmc}

The  1-form Killing spinor bilinear\footnote{There are other $\lambda$-valued 1-form bilinears which can be constructed from the Killing spinor $\e$. However,
these cannot be identified with $V$ as they are twisted 1-forms and generically are not associated with a Killing vector.}  is
\be
Z = \langle B (C \epsilon^*)^*, \Gamma_A \epsilon \rangle~ e^A
= \langle \Gamma_0 \epsilon, \Gamma_A \epsilon \rangle~ e^A~,
\la{1bl}
\ee
where $\e$ is a Killing spinor and $\langle\cdot, \cdot\rangle$ is the standard Hermitian inner product, see  \cite{sgiib}.
We require that $Z$
should be proportional to $V$, where
\be
V = -{1 \over 2} r^2 \Delta \bbe^+ + \bbe^-~.
\ee
First, evaluate $Z$ at $r=u=0$, for which $\epsilon=\eta_++ \eta_-$.
Requiring that $Z_+=0$ at $r=u=0$ forces
\be
\eta_-=0~.
\ee
Then, using $r,u$ independent $Spin(8)$ gauge transformations of the type considered in \cite{sgiib},
one can, without loss of generality, take
\be
\eta_+=p+q e_{1234}~,
\ee
for $p,q$ complex valued functions. Furthermore, on computing the component $Z_-$, one finds that
$|p|^2+|q|^2$ must be a (non-zero) constant,
or equivalently,
\bea
\label{cnorm1}
\langle \eta_+, \eta_+ \rangle = {\rm const.} \
\eea

Next, evaluate $Z_i$ at $r \neq 0$. As this component must vanish, one finds the condition
\bea
\label{hexp}
(|p|^2+|q|^2) h_\alpha
+{3 \over 4} (pq {\bar{\Phi}}_\alpha + {\bar{p}} {\bar{q}} \Phi_\alpha)
+2i (|p|^2-|q|^2) Y_{\alpha \beta}{}^\beta
\nn
-{1 \over 4} (pq {\bar{H}}_{\alpha \beta}{}^\beta + {\bar{p}} {\bar{q}} H_{\alpha \beta}{}^\beta)
+{1 \over 12} \epsilon_{\alpha \beta_1 \beta_2 \beta_3}(p^2 {\bar{H}}^{\beta_1 \beta_2 \beta_3}
+ {\bar{q}}^2 H^{\beta_2 \beta_2 \beta_3} ) = 0~,
\eea
where the 8-dimensional indices are split as $i=(\a,\bar\a)$.

It will be convenient to define
\bea
\label{et}
\tau_+ = \bigg({1 \over 4} h_i \Gamma^i +{i \over 12} Y_{\ell_1 \ell_2 \ell_3} \Gamma^{\ell_1 \ell_2 \ell_3}\bigg)
\eta_+ + \bigg({3 \over 16} \Phi_i \Gamma^i +{1 \over 96} H_{\ell_1 \ell_2 \ell_3} \Gamma^{\ell_1 \ell_2
\ell_3} \bigg) C* \eta_+~.
\eea
Then the condition ({\ref{hexp}}) implies that one can write
\bea
\label{texp}
\tau_+ = q \mu^\alpha e_\alpha -{1 \over 6}p \ {\overline{(\mu^\beta)}} \ \epsilon_{\bar{\beta}}{}^{\alpha_1 \alpha_2 \alpha_3} e_{\alpha_1 \alpha_2 \alpha_3}~,
\eea
with
\bea
\label{muexp}
(|p|^2+|q|^2) \mu^\alpha &=& {3 \over 8 \sqrt{2}} \big( ({\bar{q}})^2 \Phi^\alpha -p^2 {\bar{\Phi}}^\alpha \big)
+ \sqrt{2} i p {\bar{q}} Y^\alpha{}_\mu{}^\mu
\nonumber \\
&-&{i \over 3 \sqrt{2}}(|p|^2+|q|^2) Y_{\mu_1 \mu_2 \mu_3} \epsilon^{\mu_1 \mu_2 \mu_3 \alpha}
+{1 \over 8 \sqrt{2}}\big( ({\bar{q}})^2 H^\alpha{}_\mu{}^\mu -p^2 {\bar{H}}^\alpha{}_\mu{}^\mu \big)
\nonumber \\
&+&{\sqrt{2} \over 48}\big(-{\bar{p}}{\bar{q}} H_{\mu_1 \mu_2 \mu_3} + pq {\bar{H}}_{\mu_1 \mu_2 \mu_3} \big)
\epsilon^{\mu_1 \mu_2 \mu_3 \alpha}~.
\eea

Also, noting that
\be
\Delta = -2 r^{-2} {Z_+ \over Z_-}~,
\ee
one finds
\bea
\label{dexp1}
\Delta &=& 4 \delta_{\alpha {\bar{\beta}}} \mu^\alpha \  {\overline{(\mu^\beta)}}~,
\eea
so $\Delta \geq 0$, as expected.

\subsubsection{Algebraic conditions}

Returning to the solution of the $+$ and $-$ components of the gravitino KSE, the Killing spinor $\e$ can be written in terms
of $\eta_+$, $\tau_+$ as
\bea
\epsilon = \eta_+  + r \Gamma_- \tau_+~.
\la{ksex}
\eea
Moreover, one also finds the algebraic conditions
\bea
\label{alg1}
\big({1 \over 2} \Delta -{1 \over 8} dh_{ij}\Gamma^{ij} \big)\eta_+ +{1 \over 16} L_{ij} \Gamma^{ij} C* \eta_+
+\big({1 \over 2} h_i \Gamma^i -{i \over 6} Y_{\ell_1 \ell_2 \ell_3} \Gamma^{\ell_1 \ell_2 \ell_3}\big) \tau_+
\nn
+\big(-{3 \over 8} \Phi_i \Gamma^i +{1 \over 48} H_{\ell_1 \ell_2 \ell_3} \Gamma^{\ell_1 \ell_2 \ell_3} \bigg) C* \tau_+
=0
\eea
and
\bea
\label{alg2}
(\Delta h_i - \partial_i \Delta) \Gamma^i \eta_+
+  \big(-{1 \over 2} dh_{ij} \Gamma^{ij} +{i \over 12} X_{\ell_1 \ell_2 \ell_3 \ell_4} \Gamma^{\ell_1
\ell_2 \ell_3 \ell_4} \big) \tau_+ +{1 \over 2}  L_{ij} \Gamma^{ij} C* \tau_+ =0~.
\nn
\eea
This completes the integration of the gravitino KSEs along the  light-cone directions.

\subsection{The ${\cal S}$ components of the gravitino KSE}

Substituting the Killing spinor $\e$, (\ref{ksex}), into the gravitino KSE and evaluating the resulting expression along the directions transverse to the light cone,
  one finds

\bea
\label{skse1}
&&\tn_i \eta_+ + \bigg(-{i \over 2} \Lambda_i -{1 \over 4} h_i
-{i \over 4} Y_{i \ell_1 \ell_2} \Gamma^{\ell_1 \ell_2}
+{i \over 12} \Gamma_i{}^{\ell_1 \ell_2 \ell_3} Y_{\ell_1 \ell_2 \ell_3} \bigg) \eta_+
\cr
&&+ \bigg({1 \over 16} \Gamma_i{}^j \Phi_j -{3 \over 16} \Phi_i
-{1 \over 96} \Gamma_i{}^{\ell_1 \ell_2 \ell_3} H_{\ell_1 \ell_2 \ell_3}
+{3 \over 32} H_{i \ell_1 \ell_2} \Gamma^{\ell_1 \ell_2} \bigg) C* \eta_+=0~,
\eea

and

\bea
\label{skse2}
\tn_i \tau_+ + \bigg(-{i \over 2} \Lambda_i -{3 \over 4} h_i
+{i \over 4} Y_{i \ell_1 \ell_2} \Gamma^{\ell_1 \ell_2}
-{i \over 12} \Gamma_i{}^{\ell_1 \ell_2 \ell_3} Y_{\ell_1 \ell_2 \ell_3} \bigg) \tau_+~~~~~~~~~~~~~~~~~~~~~~~~~&&
\cr
+ \bigg(-{1 \over 16} \Gamma_i{}^j \Phi_j +{3 \over 16} \Phi_i
-{1 \over 96} \Gamma_i{}^{\ell_1 \ell_2 \ell_3} H_{\ell_1 \ell_2 \ell_3}
+{3 \over 32} H_{i \ell_1 \ell_2} \Gamma^{\ell_1 \ell_2} \bigg) C* \tau_+ ~~~~~~~~~~~&&
\cr
+ \bigg(-{1 \over 4} dh_{ij} \Gamma^j -{i \over 12} X_{i \ell_1 \ell_2 \ell_3} \Gamma^{\ell_1 \ell_2 \ell_3} \bigg) \eta_+
+ \bigg( {1 \over 32} \Gamma_i{}^{\ell_1 \ell_2} L_{\ell_1 \ell_2}-{3 \over 16} L_{ij} \Gamma^j \bigg) C* \eta_+ =0~.&&
\eea

Both the above equations are parallel transport equations along ${\cal S}$.

\subsection{The dilatino KSE}

It remains to evaluate the dilatino KSE ({\ref{akse}}) on the spinor (\ref{ksex}). A direct substitution reveals that
\bea
\label{alg3}
\bigg(-{1 \over 4} \Phi_i \Gamma^i +{1 \over 24} H_{\ell_1 \ell_2 \ell_3} \Gamma^{\ell_1 \ell_2 \ell_3}
\bigg) \eta_+ + \xi_i \Gamma^i C* \eta_+ =0~,
\eea
and
\bea
\label{alg4}
\bigg( -{1 \over 4} \Phi_i \Gamma^i -{1 \over 24} H_{\ell_1 \ell_2 \ell_3} \Gamma^{\ell_1 \ell_2 \ell_3} \bigg) \tau_+
- \xi_i \Gamma^i C* \tau_+ +{1 \over 8} L_{ij} \Gamma^{ij} \eta_+ =0~.
\eea

This concludes the evaluation of the KSEs in the IIB near-horizon geometries and their integration along the lightcone directions.

\subsection{Independent KSEs}

It is well known that the KSEs imply some of the Bianchi identities and field equations. Because of this, to find solutions, it is customary to solve the
KSEs and then impose the remaining field equations and Bianchi identities. However, we shall not do this here because of the complexity of solving the KSEs  ({\ref{alg1}}), ({\ref{alg2}}), ({\ref{skse2}}), and ({\ref{alg4}}) which contain
the $\tau$ spinor as expressed in (\ref{et}) and  (\ref{texp}). Instead, we shall show that all the KSEs which contain $\tau$ are actually implied from those  containing only  $\eta$,  i.e.~({\ref{skse1}}) and
({\ref{alg3}}), and some
of the field equations and Bianchi identities.

\subsubsection{ The ({\ref{skse2}}) condition} \la{(i)}

 The ({\ref{skse2}}) component of  KSEs is implied by ({\ref{skse1}}) and ({\ref{et}}) together with a number of
bosonic field equations and Bianchi identities. To see this, first evaluate the LHS of ({\ref{skse2}}) by substituting in
({\ref{et}}) to eliminate $\tau_+$, and use ({\ref{skse1}}) to evaluate the supercovariant derivatives of
$\eta_+$ and $C* \eta_+$.
Also evaluate
\bea
\label{auxelim1}
\bigg( {1 \over 4} {\tilde{R}}_{ij} \Gamma^j - {1 \over 2} \Gamma^j (\tn_j \tn_i - \tn_i \tn_j) \bigg) \eta_+ = 0~.
\eea
This vanishes identically, however a non-trivial identity is obtained by expanding out the supercovariant
derivative terms again using ({\ref{skse1}}).
Then, on adding ({\ref{auxelim1}}) to the LHS of ({\ref{skse2}}), with $\tau_+$ eliminated in favour of $\eta_+$
using ({\ref{et}}) and ({\ref{skse1}}) as mentioned above, one obtains, after some calculation, a term proportional to ({\ref{feq7}}).

Therefore, it follows that ({\ref{skse2}}) is implied by ({\ref{skse1}}) and ({\ref{et}}) and the bosonic field equations and Bianchi identities.
We remark that in addition to using ({\ref{feq7}}) in establishing this identity, we also make use of
({\ref{ba1}}), ({\ref{sd1}}), ({\ref{ba2}}), ({\ref{ba3}}), ({\ref{ba4}}), ({\ref{ba6}}), ({\ref{feq1}}) and ({\ref{feq3}}).

\subsubsection{The ({\ref{alg4}}) condition}
Next consider ({\ref{alg3}}) and ({\ref{alg4}}). On defining
\bea
\label{alg3b}
{\cal{A}}_1 = \bigg(-{1 \over 4} \Phi_i \Gamma^i +{1 \over 24} H_{\ell_1 \ell_2 \ell_3} \Gamma^{\ell_1 \ell_2 \ell_3}
\bigg) \eta_+ + \xi_i \Gamma^i C* \eta_+~,
\eea
and
\bea
\label{alg4b}
{\cal{A}}_2 = \bigg( -{1 \over 4} \Phi_i \Gamma^i -{1 \over 24} H_{\ell_1 \ell_2 \ell_3} \Gamma^{\ell_1 \ell_2 \ell_3} \bigg) \tau_+
- \xi_i \Gamma^i C* \tau_+ +{1 \over 8} L_{ij} \Gamma^{ij} \eta_+~,
\eea
one obtains the following identity
\bea
{\cal{A}}_2 = -{1 \over 2} \Gamma^i \tn_i {\cal{A}}_1 + \bigg( {3i \over 4} \Lambda_i \Gamma^i +{3 \over 8} h_i \Gamma^i
-{i \over 12} Y_{\ell_1 \ell_2 \ell_3} \Gamma^{\ell_1 \ell_2 \ell_3} \bigg) {\cal{A}}_1~,
\eea
where we have made use of ({\ref{skse1}}) in order to evaluate the covariant derivative in the above expression.
In addition, we also have made use of the field equations and Bianchi identities
({\ref{ba3}}), ({\ref{ba4}}), ({\ref{ba5}}), ({\ref{feq1}}), ({\ref{feq3}}) and ({\ref{feq4}}).
It follows that these conditions, together with ({\ref{alg3}}) imply ({\ref{alg4}}).

\subsubsection{The ({\ref{alg1}}) condition}

To show that ({\ref{alg1}}) is also implied as a consequence  of the  KSEs (\ref{skse1}) and (\ref{alg3}),  and the field equations and Bianchi identities,  contract ({\ref{skse2}}) with $\Gamma^i$ and use ({\ref{et}})
to rewrite the $\tau_+$ terms in terms of $\eta_+$. Then subtract
$({3 \over 16}{\bar{\Phi}}_i \Gamma^i +{1 \over 96} {\bar{H}}_{\ell_1 \ell_2 \ell_3} \Gamma^{\ell_1 \ell_2 \ell_3}) {\cal{A}}_1$
from the resulting expression
to obtain ({\ref{alg1}}). In order to obtain ({\ref{alg1}}) from these expressions, we also make use of ({\ref{ba1}}), ({\ref{sd1}}),
({\ref{ba2}}), ({\ref{feq3}}), ({\ref{ba3}}), ({\ref{feq1}}), and ({\ref{feq5}}). It follows, from section \ref{(i)} above,  that ({\ref{alg1}})
follows from the above mentioned Bianchi identities and field equations, together with ({\ref{skse1}}) and ({\ref{alg3}}).

\subsubsection{The ({\ref{alg2}}) condition}

 The ({\ref{alg2}}) condition is obtained from ({\ref{alg1}}) as follows. First
act on ({\ref{alg1}}) with the Dirac operator $\Gamma^i {\tilde{\nabla}}_i$, and use the bosonic field equations and
Bianchi identities to eliminate the $d \star_8 dh$, $dL$, $d \star_8 L$, $d \star_8 h$, $dY$, $d \star_8 Y$, $dH$ and $d \star_8 H$
terms, and rewrite $d \Phi$ in terms of $L$. Then use the algebraic conditions ({\ref{alg3}}) and ({\ref{alg4}}) to eliminate the
$\xi$-terms from the resulting expression. The terms involving $\Lambda$ then vanish as a consequence of ({\ref{alg1}}).

Next consider the $dh$-terms; after some calculation, these can be rewritten
as
\bea
 {1 \over 2} dh_{ij} \Gamma^{ij} \tau_+ -{7 \over 32} h_\ell \Gamma^\ell dh_{ij} \Gamma^{ij} \eta_+
 + \big(-{1 \over 64} \Phi_\ell \Gamma^\ell +{1 \over 384} H_{\ell_1 \ell_2 \ell_3} \Gamma^{\ell_1 \ell_2 \ell_3}
 \big) dh_{ij} \Gamma^{ij} C*\eta_+ \ .
 \nonumber
 \eea
The $dh$ terms involving $\eta_+$ and $C* \eta_+$ in the above expression are then eliminated, using
({\ref{alg1}}).
On collating the remaining terms, one finds that those involving $\Delta$ (but not $d \Delta$) are
\bea
-\Delta h_j \Gamma^j \eta_+ \ .
\eea
It is also straightforward to note that the terms involving ${\bar{L}}$ vanish,
whereas the terms involving $X$ and $L$ can be rewritten as
\bea
-{1 \over 2} L_{ij} \Gamma^{ij} C* \tau_+ -{i \over 12} X_{\ell_1 \ell_2 \ell_3 \ell_4} \Gamma^{\ell_1 \ell_2 \ell_3 \ell_4} \tau_+~,
\eea
where the anti-self-duality of $X$ has been used to simplify the expression.
The remaining terms which are linear in $\tau_+, C*\tau_+$ and quadratic in $h, Y, \Phi, {\bar{\Phi}}, H, {\bar{H}}$ can be shown to vanish after some computation.
After performing these calculations, the condition which is obtained is ({\ref{alg2}}).

\setcounter{equation}{0}
\section{Solution of  KSEs and geometry of horizons}

\subsection{The linear system} \label{systsec}

It is straightforward to derive the linear system associated with the ({\ref{skse1}}) and (\ref{alg3}) KSEs  evaluated on the spinor
\bea
\eta_+ = p 1 + q e_{1234}~,
\eea
where we have used a local $Spin(8)\cdot U(1)$ transformation to arrange such that $p$, $q$ are real functions on ${\cal S}$, and as required from the bilinear matching condition
 take
\bea
p^2+q^2=1~.
\eea
Substituting this spinor into  ({\ref{skse1}}) and (\ref{alg3}), one obtains a  linear system which is explicitly given in appendix A. This can be used to express some of the fluxes in terms of geometry and
find the conditions on the geometry of the horizon sections ${\cal S}$ imposed by supersymmetry.

There are three different types of backgrounds which arise naturally in the investigation of the solutions of the linear system in appendix A:

\begin{itemize}

\item The generic horizons for which the Killing spinor is chosen such that $p^2+q^2=1$,  $p^2-q^2\not=0$, and $p$ and $q$ not identically zero at the horizon though they may vanish at some points.

\item The $Spin(7)$ horizons for which $p^2+q^2=1$ and $p=q$.

\item The pure $SU(4)$ horizons for which either $p=1, q=0$ or $p=0, q=1$.

\end{itemize}

 Clearly, the generic horizons include both the $Spin(7)$ and pure $SU(4)$ horizons. Typically, as the Killing spinor is parallel transported along the horizon section, it will change type as the only requirement is that $p^2+q^2=1$. The solution of the linear system for all three cases is given in appendix A. In what follows, we shall explore the consequences which arise on the topology and
 geometry of horizon sections ${\cal S}$ from the solution of the linear system and hence from the solution of the KSEs.

\subsection{Generic horizons} \label{systsec1}

\subsubsection{ Solution of the linear system}

The general solution
of the linear system  is described in appendix A under the assumption that the functions $p$ and $q$ may vanish at isolated points.
The main properties of the solution are the following. All complex fluxes $\xi, \Phi$ and $H$ can be expressed in terms of the real fluxes $Y$ and
 the geometry $\Omega$. Moreover, the solution of the linear system does {\it not} contain  a condition which involves only the spin connection of ${\cal S}$.

\subsubsection{Topology and geometry of ${\cal S}$}

The gravitino KSE as reduced to ${\cal S}$ in ({\ref{skse1}}) is a parallel transport equation of sections of a $\lambda^{{1\over2}}\otimes \Sigma^+$
bundle over ${\cal S}$, where $\lambda^{{1\over2}}\otimes \Sigma^+$ is a $Spin_c(8)=Spin(8)\cdot U(1)$ bundle and $\Sigma^+$ is the complexified spin bundle over ${\cal S}$ associated with the positive chirality Majorana-Weyl representation of $Spin(8)$.
The real rank of $\lambda^{{1\over2}}\otimes \Sigma^+$ is 16. An application of \cite{gpdt} implies that the supercovariant connection as
restricted on ${\cal S}$ given in ({\ref{skse1}}) has holonomy contained in $SL(16, \bR)$. The existence of a solution to the gravitino KSE requires that the holonomy of the supercovariant connection reduces to $SL(15, \bR)$, and so
$\lambda^{{1\over2}}\otimes \Sigma^+$ must admit a nowhere vanishing section spanned by the parallel spinor. The existence of nowhere vanishing  sections typically impose a topological restriction
on the spin bundle and so on a manifold. However, a priori this may not be   the case here as $\lambda^{{1\over2}}\otimes \Sigma^+$ has rank much bigger than the dimension of ${\cal S}$
 and so it always admits
such a nowhere vanishing section.

Nevertheless, the existence of a solution to the KSEs requires that the structure group of ${\cal S}$ reduces form $SO(8)$ to a subgroup of $U(4)$. To see this, recall that ${\cal S}$ admits a nowhere vanishing spinor $\eta_+=p1+q e_{1234}$ which is a section of $\lambda^{{1\over2}}\otimes \Sigma^+$. One can consider the bilinears of $\eta_+$. There are two kinds of bilinears which can be constructed from $\eta_+$.  First the bilinears constructed from $\eta_+$ and itself with respect to the standard Hermitian inner product of $Spin(8)$.  These bilinears are forms on ${\cal S}$.
In addition one can also consider the bilinears constructed from $\eta_+$ and  its complex conjugate $C*\eta_+$.  Such bilinears are not forms on ${\cal S}$. Instead they are forms twisted by $\lambda$ as they carry a $U(1)$ charge.

Consider first the bilinears of $\eta_+=p1+q e_{1234}$ which are forms on ${\cal S}$. We find that a basis in the ring  of these bilinears is
\bea
\omega=-i (p^2+q^2) \delta_{\a\bar\b} e^\a\wedge e^{\bar\b}~,~~~\psi=\, pq\, \mathrm {Re}\, (e^1\wedge e^2\wedge e^3\wedge e^4)~,
\la{bbb}
\eea
with $p^2+q^2=1$. Since $\omega$ vanishes nowhere,  it is an almost Hermitian form on ${\cal S}$. In addition $\psi$ defines a (4,0)-form on ${\cal S}$.  But as $p$ and $q$ can vanish
at some points on ${\cal S}$, $\psi$   is not nowhere
vanishing. As a result, the structure group of  ${\cal S}$ reduces from $SO(8)$ to a subgroup of $U(4)$ rather than to a subgroup of $SU(4)$. ${\cal S}$ is an almost
Hermitian 8-dimensional manifold.

Next, the form bilinears of $\eta_+=p1+q e_{1234}$ which are twisted by $\lambda$ are
\bea
\rho=-i pq \delta_{\a\bar\b} e^\a\wedge e^{\bar\b}~,~~~\tau= p^2\, e^1\wedge e^2\wedge e^3\wedge e^4 +q^2\, e^{\bar 1}\wedge e^{\bar  2}\wedge e^{\bar3}\wedge e^{\bar 4}~.
\la{tbi}
\eea
As a result $\lambda\otimes (\Lambda^{4,0}({\cal S})\oplus \Lambda^{0,4}({\cal S}))$ admits a nowhere vanishing section.

Furthermore, as has been mentioned,  the KSEs  solved in appendix A do not impose any additional
geometric restrictions on ${\cal S}$. As a result, the horizon sections ${\cal S}$ are  $\mathrm {Spin}_c$ almost Hermitian 8-dimensional manifolds.
The only remaining potential
 restrictions on ${\cal S}$ are those imposed by the field equations and Bianchi identities which we shall investigate in some special cases below.

So far, we have performed our analysis without distinguishing whether the IIB scalars take values in the upper half plane or in the fundamental domain.
Now we return to address this issue. If the IIB scalars take values in the upper half plane, then the line bundle $\lambda$ is topologically trivial
as it is the pull back of a trivial bundle over the upper half plane with respect to a smooth map. Nevertheless, if the IIB scalars are non-constant\footnote{It is not apparent that
there are smooth solutions for which ${\cal S}$ is compact and the IIB scalars are non-trivial. }  $\lambda$ will be geometrically twisted
as it will have non-vanishing curvature.

In the case that the IIB scalars take values in the fundamental domain, $\lambda$ may not be topologically trivial. This arises from
the analysis of stringy cosmic  string solutions in \cite{vafa} where there are solutions which are smooth and the horizon section
of the cosmic strings is a 2-sphere, e.g.~the configuration of 24 cosmic strings. Of course, our case here is more general and ${\cal S}$ is not
a complex manifold. Nevertheless, it is an indication that $\lambda$ over ${\cal S}$ may not be topologically trivial in the fundamental domain case.

\subsection{$Spin(7)$ Horizons}

\subsubsection{Solution of the linear system}

The spinor $\eta_+$ can be chosen as $\eta_+={1\over\sqrt2}(1+e_{1234})$.
The solution of the linear system is given as in appendix A after setting $p=q=1/\sqrt2$ in the solution of the generic case.
Furthermore, the expression for  $\mu$ can be simplified as
\bea
{\mu_{\bar\a}\over\sqrt2}=-{i\over2} \Lambda_{\bar\a}+i Y_{\bar\a\l}{}^\l
+ {i\over 3} \e_{\bar\a}{}^{\l_1\l_2\l_3}
Y_{\l_1\l_2\l_3} ~,
\eea
which is consistent with our previous results on IIB horizons with only 5-form flux.

\subsubsection{Topology and geometry of horizon sections}

The analysis of the consequences for the existence of a parallel $\eta_+$ spinor on the holonomy of the supercovariant connection
as reduced on ${\cal S}$ is similar to the generic case.  To find whether the structure group of ${\cal S}$ reduces to a subgroup
of $SO(8)$, we again compute the form bilinears associated with $\eta_+$.  In this case, there is a single nowhere vanishing
4-form bilinear\footnote{The almost hermitian form $\omega$ and the (4,0)-form  $\chi$ may not be globally defined on ${\cal S}$ but $\phi$ is. Observe that we have normalized
$\chi$ differently in \cite{sgiib}.}
\bea
\phi=-{1\over 2}\omega\wedge \omega+ 4\,{\mathrm{Re}} \chi
\eea
which is a representative of the fundamental form of $Spin(7)$. Therefore the structure group of ${\cal S}$ reduces to $Spin(7)$, where  $\chi={1\over 4!} \e_{\a_1\a_2\a_3\a_4} e^{\a_1}\wedge e^{\a_2}\wedge e^{\a_3}\wedge e^{\a_4}$.
It is known that  the structure group of an 8-dimensional compact  spin manifold reduces from $SO(8)$ to  $Spin(7)$ provided that \cite{isham}
\bea
\pm e-{1\over2} p_2+ {1\over8} p_1^2=0~,
\la{topob}
\eea
where $e$ is the Euler class, and $p_2$ and $p_1$ are the second and first Pontryagin classes, respectively. This condition restricts the topology
of ${\cal S}$.  Furthermore, a direct inspection of the solution of the linear system reveals that there are no additional geometric
conditions on the $Spin(7)$ structure of ${\cal S}$. Thus ${\cal S}$ can be any manifold with a $Spin(7)$ structure.

\subsection {Pure $SU(4)$  horizons} \label{systsec2}

\subsubsection{Solution of the linear system}

The $\eta_+$ spinor in this case can be chosen as $\eta_+= 1$.  The solution of the linear system has been given in appendix A.
In particular, many components of the complex fields, like $H$, can be expressed
in terms of the components of the 5-form and components of the spin connection of spacetime. Again, the solution of the linear system in appendix
A does not yield a condition which restricts the spin connection of ${\cal S}$.

\subsubsection{Topology and geometry of horizon sections}

A similar analysis to that presented for the generic case reveals that the structure group of ${\cal S}$ reduces to a subgroup of $U(4)$.
In this case, it is instructive to also consider the mixed bilinears of $\eta_+$ with $\tilde \eta_+=C*\eta_+=e_{1234}$ in (\ref{tbi}). In addition to the almost
hermitian form $\omega$ which arises as a bilinear of $\eta_+$, as in (\ref{bbb}), there is a  $\lambda$-twisted (4,0)-form   bilinear given locally by  $\tau$ for $p=1, q=0$.  Therefore
the line bundle $\lambda\otimes \Lambda^{4,0}$ on ${\cal S}$ admits a nowhere vanishing section, and so it is topologically trivial. Thus $\lambda$
can be identified with the anti-canonical bundle of ${\cal S}$. Furthermore, the solution of the linear system in appendix A does not reveal any additional geometric constraints on ${\cal S}$.  As a result, ${\cal S}$
is any almost Hermitian spin${}_c$ manifold.

Next let us turn to examine the upper half plane and fundamental domain cases. As in the generic class, if the IIB scalars take values in the upper
half plane, $\lambda$ is topologically trivial.  The structure group of ${\cal S}$ then reduces to a subgroup of $SU(4)$. However, $\lambda$ can be geometrically non-trivial
as its curvature may not vanish provided that the IIB scalars are non-trivial functions\footnote{It is not apparent that there exist
smooth solutions with ${\cal S}$ compact and non-trivial scalars.}.  On the other hand, if the IIB scalars take values in the fundamental domain, as has been explained
in the generic case, $\lambda$ may not be topologically trivial on ${\cal S}$.

\newsection{Special cases}

\subsection{Horizons with only 5-form flux}

It is well known that IIB supergravity has two consistent truncations. One truncation leads to a sector with
only 5-form fluxes active, while for the other truncation the
5-form field strength vanishes and both the 3-form and 1-form field strengths are real. The latter is the common sector. In both cases, the near-horizon
geometries are compatible with both these truncations. It is straightforward to see this in the 5-form sector. The near-horizon geometries with only 5-form fluxes
can be recovered from the general IIB horizons that we have investigated by setting
\bea
P=G=0~.
\eea
Conversely, the 5-form horizons investigated in \cite{iibhorf5} can be embedded into general IIB horizons exhibiting only non-vanishing 5-form fluxes.
As a result, all the examples of IIB horizons constructed in \cite{iibhorf5} can be embedded into the general IIB horizons.

\subsection{Common sector horizons}

The embedding of the common sector horizons into general IIB horizons is not as straightforward as that described in the previous section
for horizons with only 5-form fluxes. To describe the embedding of common sector horizons, let us recall how the common sector
is obtained from the general IIB theory.  For this set $F=0$,  and as has already been mentioned, choose both $P$ and $G$ to be real.
Observe that $Q=0$, and the Bianchi identities of IIB supergravity can be solved after setting
\bea
P={1\over2} d\phi~,~~~G=-e^{-{1\over2}\phi} {\cal H}~,
\la{formred}
\eea
provided that  $d {\cal H}=0$. These field definitions yield the common sector in the
string frame provided that in addition we relate the IIB metric, $ds^2_{\mathrm{IIB}}$, with the common sector metric, $ds^2_{\mathrm{CS}}$, as
\bea
ds^2_{\mathrm{IIB}}= e^{-{1\over2}\phi} ds^2_{\mathrm{CS}}~,
\la{metrred}
\eea
and identify $\phi$ and ${\cal H}$ with the dilaton and  the NS-NS 3-form field strength, respectively.

Furthermore,  the KSEs of the common sector are obtained from those of  IIB  provided that the IIB supersymmetry parameter is related to that of the common sector
as
\bea
\epsilon_{\mathrm{CS}}= e^{{1\over8}\phi} \epsilon_{\mathrm{IIB}}~.
\eea
In particular, we find that the KSEs of the common sector are
\bea
\nabla_M^{\pm}\epsilon_{\mathrm{CS}}^\pm&=&0~,
\cr
(\Gamma^M\partial_M\phi\mp {1\over12} \Gamma^{MNR} {\cal H}_{MNR})\epsilon_{\mathrm{CS}}^\pm&=&0~,
\eea
where $\nabla^{\pm}=\nabla\pm{1\over2}{\cal H} $, $\nabla$ the Levi-Civita connection of the common sector metric,   and $C*\epsilon_{\mathrm{CS}}^\pm=\pm \epsilon_{\mathrm{CS}}^\pm$.

It remains to show how the common sector near-horizons  are embedded into general IIB horizons. For this, one has to demonstrate that after making the
 field redefinitions (\ref{formred}) and (\ref{metrred}), one can adapt a coordinate system such that the common sector near-horizon fields give rise to IIB near-horizon fields. Indeed consider first the metric. If the common sector  metric is in near-horizon form, the IIB metric can also be put in near-horizon form provided that we make a coordinate transformation
\bea
r_{\mathrm{IIB}}= e^{-{1\over2} \phi} r_{\mathrm{CS}}~,~~
\la{corred}
\eea
the 1-form $h_{\mathrm{CS}}$ is replaced with $h_{\mathrm{IIB}}= h_{\mathrm{CS}}+{1\over2} d\phi$, and the metric of the horizon section ${\cal S}$
is conformally scaled as $g_{\mathrm{IIB}}=e^{-{1\over2}\phi} g_{\mathrm{CS}}$.

Next turn to the 3-form field strength. Suppose that   ${\cal H}$ is  in a near-horizon form. $G$ in (\ref{formred}) can also be put into a near-horizon form as well since the conformal factor in the definition of $G$ is absorbed in the coordinate transformation for $r$ in (\ref{corred}) and the redefinition of $h$. One also has
to re-scale $H$ as $H_{\mathrm{IIB}}=e^{-{1\over2}\phi} H_{\mathrm{CS}}$. Observe now that $H_{\mathrm{CS}}$ is closed as expected. To make connection with our results for heterotic horizons, one can compensate the overall sign in the definition of ${\cal H}$ by changing the sign of $L$, $\Phi$ and $H$.

A consequence of our analysis above is that we can embed all the near-horizon geometries we have found in heterotic theory into IIB. We remark that all the horizons we had investigated in \cite{hhor}
have $d{\cal H}=0$ and so they can be thought as solutions of the common sector as well. In this way, all the explicit heterotic near-horizon  geometries found in \cite{hhor} can be
utilized to give explicit examples of IIB horizons.  Note also that all the IIB horizons with non-trivial fluxes that arise from the embedding of
heterotic horizons in IIB preserve more than two supersymmetries and admit additional isometries.

\newsection{Pure $SU(4)$ horizons with constant scalars}

\subsection{Solution to the linear system}

A special class of horizons are those for which the IIB scalars are constant.  In this case, there is a significant simplification of
the solution to the linear system because $P=Q=0$.

The solution to the KSEs given in appendix A simplifies somewhat. In particular, one finds that the fluxes of pure $SU(4)$ horizons with constant scalars can be expressed\footnote{If $v$ is  $\ell$-vector and $\a$ a k-form, then $(v\cdot \a)_{i_1\dots i_{k-\ell}}= v^{j_1\dots j_\ell} \a_{j_1\dots j_\ell i_1\dots i_{k-\ell}}$.} as
\bea
&&Y={1\over4} (d\omega-\theta_{\mathrm{Re}\chi} \wedge \omega)+ \mathrm{Im} \Phi \cdot \mathrm{Re} \chi~,~~~\omega\cdot H^{1,2}=H^{0,3}=\Phi^{0,1}=0~,
\cr
&&H^{3,0}=(\theta_{\mathrm{Re}\chi}-\theta_\omega)\cdot \chi~,~~~H_{\bar\a\l_1\l_2}= {1\over2} {\cal N}_{\bar\nu_1\bar\nu_2\bar\a} \e^{\bar\nu_1\bar\nu_2}{}_{\l_1\l_2}-2 \delta_{\bar\a[\l_1} \Phi_{\l_2]}~,~~~h=\theta_{\mathrm{Re}\chi}~,
\nonumber \\
\la{constscalars}
\eea
where $\theta_\omega$ and $\theta_{\mathrm{Re}\chi}$ are the Lee forms\footnote{We define $(\theta_\omega)_i=-I^j{}_i \nabla^k\omega_{kj}$ and $(\theta_{\mathrm{Re}\chi})_i={2\over3} \mathrm{Re}\, \chi_i{}^{j_1j_2j_3} \nabla^k \mathrm{Re}\,\chi_{kj_1j_2j_3}$.} of the almost Hermitian form $\omega$ and the (4,0) form $\chi={1\over 4!} \e_{\a_1\a_2\a_3\a_4} e^{\a_1}\wedge e^{\a_2}\wedge e^{\a_3}\wedge e^{\a_4}$
 of the $SU(4)$ structure, respectively, and
 \bea
 {\cal N}^i_{jk} =I^m{}_j\nabla_m I^i{}_k-I^m{}_k\nabla_m I^i{}_j- I^i{}_m (\nabla_j I^m{}_k-\nabla_k I^m{}_j)~,
 \eea
 is the Nijenhuis tensor of the almost complex structure $I$.
The fluxes $\Phi^{1,0}$ and the traceless part of $H^{1,2}$ are not restricted by the KSEs.

\subsection{Topology and geometry of horizons}

As the scalars are constant, the line bundle $\lambda$ is trivial.  As a consequence, the bilinears of both $\eta_+$ and $\tilde\eta_+=e_{1234}$ are forms on ${\cal S}$
and in particular $\chi$ is a nowhere vanishing (4,0)-form on ${\cal S}$.  Thus the structure group of ${\cal S}$ reduces to a subgroup of $SU(4)$.

As in the $Spin(7)$ case, there are topological obstructions to reduce the structure group of ${\cal S}$ from $SO(8)$ to $SU(4)$. Since $Spin(7)$ and $SU(4)$ have the same maximal torus\footnote{We thank Simon Salamon for discussions on this.}
the obstruction (\ref{topob}) is also an obstruction in the $SU(4)$ case. In particular using the relation between the Pontryagin and Chern  classes $p_1=c_1^2-2 c_2$ and $p_2=c_2^2-2 c_1 c_3+ 2 c_4$
on almost complex manifolds, see e.g.~\cite{milnor},  and the identification of the Euler class $e=c_4$, (\ref{topob}) is satisfied provided $c_1=0$ which vanishes as a consequence of $SU(4)$ structure.
The orientation induced by the almost complex structure leads to the choice of the plus sign in (\ref{topob}).
There may be additional obstructions to reduce the structure group from $SO(8)$ to $SU(4)$  which take values in cohomology with $\bZ_k$ coefficients.

The solution of the linear system  does not impose any additional conditions on the geometry of ${\cal S}$. Therefore
${\cal S}$ is an almost Hermitian manifold with an $SU(4)$ structure.

\subsection{Magnetic 3-form flux deformations}

\subsubsection{Complex deformation of horizons with 5-form fluxes}

A class of horizons with 3-form flux can be constructed as a deformation of horizons with 5-form fluxes. Recall that the horizons with only 5-form fluxes
are 2-SCYT manifolds. The  deformation we shall investigate here  lifts the 2-strong condition but  the horizon  remains a complex manifold.
For this, we take\footnote{Note that since $\Phi = 0$ it follows from (\ref{puremu}), using (\ref{puresol}), that $\tau_+ = 0$ and therefore the Killing spinor in this case is just $\epsilon= \eta_+$, which is a considerable simplification.}
\bea
\Phi=H^{0,3}=H^{3,0}=H^{2,1}=0~,~~~Y^{0,3}=Y^{3,0}=0~.
\eea
In this case, the conditions stated in appendix A imply that ${\cal S}$ is a Hermitian manifold and in addition
\bea
Y={1\over4} (d\omega-\omega\wedge \theta_\omega)~,~~~\omega\cdot H^{1,2}=0~,~~~\theta_\omega=\theta_{\mathrm{Re}\chi}~,~~~h=\theta_\omega~,
\eea
where $\theta_\omega$ and $\theta_{\mathrm{Re}\chi}$ are the Lee forms of the Hermitian form $\omega$ and $\mathrm{Re}\chi$ is the
real part of the (4,0)-form $\chi$. Observe that the traceless part of $H^{1,2}$ is not restricted by the KSEs. Since ${\cal S}$ is a Hermitian manifold with an $SU(4)$ structure, the  condition $\theta_\omega=\theta_{\mathrm{Re}\chi}$ implies that ${\cal S}$ admits  a connection $\hat\nabla=\nabla+{1\over2} \mathring H$ with skew-symmetric torsion $ \mathring H$ which has holonomy $SU(4)$,
i.e.~${\rm hol}(\hat\nabla)\subseteq SU(4)$, where $ \mathring H$ is uniquely determined in terms
of the metric $g$ and complex structure $I$ of the horizon section as $ \mathring H=-i_I d\omega$. Thus ${\cal S}$ admits a CYT structure.

It remains to solve the field equations and Bianchi identities. One can show after a long but straightforward  calculation  that
all these are satisfied provided that
\bea
d(\omega\wedge \mathring H )={i\over2} H\wedge \bar H~,~~~dH=0~,~~~
h^{\bar\g} H_{\bar\g\a\bar\b}+2i  H_{\a \bar\g_1\bar\g_2} Y_{\bar\b}{}^{\bar\g_1\bar\g_2}=0~.
\la{puregeom}
\eea
The first condition is a generalization of the 2-strong condition found in \cite{iibhorf5} for horizons with only
5-form fluxes. In particular, the 3-form flux appears as a source term in the rhs of the 2-strong condition.
Observe also that the last condition together with $\omega\cdot H^{1,2}=0$ imply that $H\cdot Y=0$ and so $H$ and $Y$ must be orthogonal.

\subsubsection{Examples}

To construct examples, one begins with an 8-dimensional CYT manifold and then  imposes  the conditions (\ref{puregeom}). Examples of CYT manifolds can be constructed
 as toric fibrations over K\"ahler manifolds \cite{cyt1, cyt2, cyt3}. This construction has been generalized in \cite{iibhorf5}  and adapted to solve the 2-SCYT condition of IIB horizons with only 5-form fluxes. Here we shall use the analysis of \cite{iibhorf5} to find geometries that solve (\ref{puregeom}).

First consider ${\cal S}$ as a $T^2$ fibration over a 6-dimensional K\"ahler manifold $X^6$ with K\"ahler form $\omega_{(6)}$. Let $(\lambda^1, \lambda^2)$ be a principal bundle connection on ${\cal S}$. In order for ${\cal S}$ to admit an $SU(4)$ structure compatible with a connection with skew-symmetric torsion, the curvature ${\cal F}^1=d\lambda^1$ must be identified with the Ricci
form of $X^6$,
i.e.
\bea
\rho_{(6)}=-{\cal F}^1~,
\eea
 and ${\cal F}^2=d\lambda^2$ must be (1,1) and traceless.  In the latter case, such connections always exist on complex line bundles.
The metric and Hermitian form on ${\cal S}$ is chosen as
\bea
ds^2={2\over k} [(\lambda^1)^2+ (\lambda^2)^2]+ ds^2(X^6)~,~~~\omega_{(8)}=-{2\over k}\lambda^1\wedge \lambda^2+\omega_{(6)}~,
\eea
where ${\cal F}^1_{ij} \omega_{(6)}^{ij}=k$ and $k$ is required  to be constant\footnote{Since $X^6$ is K\"ahler, observe that $k$ is constant
provided ${\cal F}^1$ is co-closed. Equivalently $X^6$ has positive constant scalar curvature. }.

 This choice of Hermitian structure specifies both $\mathring H$ and $Y$. It remain to choose $H$.  For this write
 \bea
 H=\bar\tau\wedge \a+\b~,~~~\tau={1\over\sqrt 2} (\lambda^1+i\lambda^2)~,
 \eea
 where $\a$ and $\b$ are complex (1,1)- and (1,2)-forms on $X^6$. The requirement that $H$ is traceless implies that
 \bea
 \omega_{(6)}\cdot \a=\omega_{(6)}\cdot \b=0~.
 \eea
 Furthermore $dH=0$ implies that
 \bea
 d\a=0~,~~~d\b+{1\over\sqrt 2}\a\wedge [{\cal F}^1-i{\cal F}^2]=0~.
 \eea
 Next the first equation in (\ref{puregeom}) is solved provided
 \bea
 \a\wedge \bar\a=0~,~~~\a\wedge \bar\b=0~,~~~\omega_{(6)}\wedge [({\cal F}^1)^2+({\cal F}^2)^2]={i k\over4} \b\wedge \bar\b~.
 \la{purex1}
 \eea
 The simplest case is to take $\a=0$. Then since $\theta_\omega=-\lambda^2$ and
 \bea
 Y={1\over 2k}(\lambda^1\wedge {\cal F}^2-\lambda^2\wedge {\cal F}^1)+{1\over4} \lambda^2\wedge \omega_{(6)}
 \eea
 the last condition in (\ref{puregeom}) is also satisfied.

 Therefore to find examples of near-horizon geometries, one must find a K\"ahler manifold $X^6$ with positive constant scalar curvature that admits a traceless (1,2) closed form $\b$
 that satisfies the last equation in (\ref{purex1}).

 To find solutions, one possibility is to take $X^6$ to either be a K\"ahler-Einstein manifold or products of K\"ahler-Einstein manifolds. For example, one can take $X^6=S^2\times S^2\times T^2$.
 Normalizing the metric of $S^2$'s such that ${\cal F}^1=\omega^1+\omega^2$ where $\omega^1$ and $\omega^2$ are the K\"ahler forms on the $S^2$'s, $k=4$, and choosing
 \bea
 \omega_{(6)}&=&-d\varphi^1\wedge d\varphi^2+\omega^1+\omega^2~,~~~ H= m (d\varphi^1-id\varphi^2)\wedge (\omega^1-\omega^2)~,~~~
 \cr
 {\cal F}^2&=&\ell (\omega^1-\omega^2)~,~~~\ell\in \bZ~,
 \eea
 all equations are solved provided that $2 |m|^2=\ell^2-1$, where $\varphi^1$ and $\varphi^2$ are the angular coordinates of $T^2$ in $X^6$.
 This generalizes the examples of 2-SCYT near-horizon geometries with only 5-form flux in \cite{iibhorf5}.

\newsection{Concluding remarks}

We have solved the KSEs for all IIB horizons that admit at least one supersymmetry. This has been done by first integrating the KSEs along the lightcone
directions and then identifying  the independent equations using the Bianchi identities, field equations and bilinear matching condition.
Then the independent KSEs are solved using spinorial geometry which leads to three different cases, the generic horizons, the $Spin(7)$ horizons and the pure $SU(4)$ horizons.
We have found that the requirement of IIB horizons to admit one supersymmetry puts rather weak geometric restrictions on the horizon sections ${\cal S}$.  In particular
for generic horizons and  pure $SU(4)$ horizons, the horizon sections can be any 8-dimensional almost Hermitian spin${}_c$ manifold.   A similar result also applies for $Spin(7)$ horizons. We have also described some topological aspects of the horizon sections and how the various fluxes are expressed in terms of the geometry.

We have also explained how horizons with only 5-form fluxes and common sector horizons, which had been investigated previously, are included
in our analysis. As a result all the examples constructed in these two special cases can be embedded in the full IIB theory. Furthermore, we give some examples for which, in addition to the KSEs,  we also solve the field equations and Bianchi identities. In particular, we focus on horizons with constant scalars which have complex horizon sections, and we find
a generalization of the 2-SCYT structure which had appeared for horizons with 5-form fluxes only.

The construction of examples with non-trivial scalars  taking values on the upper half-plane after a $SL(2,\bR)$ identification is natural within the context of 10-dimensional type of F-theory \cite{vafa2}. In particular, one may consider
$T^2$-fibrations over ${\cal S}$.  However there are some differences. As we have mentioned, ${\cal S}$ are almost complex manifolds instead of K\"ahler, which mostly arise in the
context of F-theory. In addition, the example of horizons with constant scalars
that we have explicitly constructed indicates that  ${\cal S}$ is  not K\"ahler because of the presence of form fluxes.  Nevertheless in the complex case,  it may be possible to construct examples imitating techniques that have been
employed in F-theory.

In most of our considerations, like the solution of the KSEs and the description of the geometry of the horizon sections, we have not used the compactness of ${\cal S}$. So our results apply to both black holes and brane horizons.
For applications to black holes, it would be of interest to enforce compactness of ${\cal S}$. This has been done for M-horizons in \cite{mindex}, and after an application of the index theorem for
 the Dirac operator, it has led to the conclusion
that all M-horizons preserve an even number of supersymmetries and admit an $\mathfrak{sl}(2,\bR)$ symmetry. A similar application may be possible in IIB. However, there are some differences between M-theory and IIB. One of them is that unlike the M-horizon sections which are odd dimensional,  the IIB near-horizon sections are even dimensional and so  the index of the Dirac operator is not expected to vanish. As the vanishing of the Dirac index has been instrumental in proving supersymmetry enhancement for M-horizons, a similar application in IIB will require some modification.
Nevertheless, it is expected that even if one cannot prove supersymmetry enhancement for IIB black hole horizons, it may be possible to relate the number of supersymmetries
preserved  in terms of the index of a Dirac operator on the horizon sections. Such a relation will generalize the classic formula $N=\mathrm{index} (D)$ which
relates the number of parallel spinors $N$ on irreducible holonomy $Spin(7)$,  $SU(4)$ and $Sp(2)$ 8-dimensional manifolds, for $N=1,2$ and $3$ respectively, to the index of the Dirac operator.
In turn such a formula will provide a topological criterion for 8-dimensional manifolds to admit Killing spinors.

\vskip 0.5cm
\noindent{\bf Acknowledgements} \vskip 0.1cm
\noindent  UG is supported by the Knut and Alice Wallenberg Foundation.
JG is supported by the STFC grant, ST/1004874/1.
GP is partially supported by the  STFC rolling grant ST/J002798/1.
\vskip 0.5cm

\setcounter{section}{0}\setcounter{equation}{0}

\appendix{Solution of the linear system}

\setcounter{subsection}{0}

\subsection{The linear system}

It is straightforward to derive the linear system associated with the KSE ({\ref{skse1}}) evaluated on the spinor
$\eta_+ = p 1 + q e_{1234}$,
where $p$, $q$ are real functions on ${\cal S}$ and $p^2+q^2=1$.
Substituting this spinor into  ({\ref{skse1}}), one obtains,

\bea
\label{lin1}
\partial_\alpha p + p \big( -{i \over 2} \Lambda_\alpha -{1 \over 4} h_\alpha +{1 \over 2} \Omega_{\alpha, \mu}{}^\mu
-i Y_{\alpha \mu}{}^\mu \big) + q \big({1 \over 4} H_{\alpha \mu}{}^\mu -{1 \over 4} \Phi_\alpha \big) =0~,
\eea
\bea
\label{lin2}
p \big( {1 \over 2} \Omega_{\alpha, {\bar{\mu}}_1 {\bar{\mu}}_2} -i Y_{\alpha {\bar{\mu}}_1 {\bar{\mu}}_2}
+ i \delta_{\alpha [{\bar{\mu}}_1} Y_{{\bar{\mu}}_2 ] \lambda}{}^\lambda +{1 \over 24} \delta_{\alpha [{\bar{\mu}}_1}
\epsilon_{{\bar{\mu}}_2]}{}^{\lambda_1 \lambda_2 \lambda_3} H_{\lambda_1 \lambda_2 \lambda_3} \big)
\nonumber \\
+q \big(-{1 \over 4} \Omega_{\alpha , \lambda_1 \lambda_2} \epsilon^{\lambda_1 \lambda_2}{}_{{\bar{\mu}}_1 {\bar{\mu}}_2}
+{1 \over 4} H_{\alpha {\bar{\mu}}_1 {\bar{\mu}}_2} -{1 \over 8} \delta_{\alpha [{\bar{\mu}}_1}
H_{{\bar{\mu}}_2] \lambda}{}^\lambda +{1 \over 8} \delta_{\alpha [{\bar{\mu}}_1} \Phi_{{\bar{\mu}}_2]} \big) =0~,
\eea
\bea
\label{lin3}
\partial_\alpha q + q \big( -{i \over 2} \Lambda_\alpha -{1 \over 4} h_\alpha -{1 \over 2} \Omega_{\alpha, \mu}{}^\mu
+{1 \over 24} H_{ {\bar{\lambda}}_1 {\bar{\lambda}}_2 {\bar{\lambda}}_3} \epsilon^{  {\bar{\lambda}}_1 {\bar{\lambda}}_2 {\bar{\lambda}}_3}{}_\alpha \big)
\nonumber \\
+ p \big(-{1 \over 8} H_{\alpha \mu}{}^\mu -{i \over 3}  Y_{ {\bar{\lambda}}_1 {\bar{\lambda}}_2 {\bar{\lambda}}_3} \epsilon^{  {\bar{\lambda}}_1 {\bar{\lambda}}_2 {\bar{\lambda}}_3}{}_\alpha
-{1 \over 8} \Phi_\alpha \big) =0~,
\eea
\bea
\label{lin4}
\partial_{\bar{\alpha}} p + p \big(-{i \over 2} \Lambda_{\bar{\alpha}} -{1 \over 4} h_{\bar{\alpha}} +{1 \over 2} \Omega_{{\bar{\alpha}}, \mu}{}^\mu +{1 \over 24} H_{\lambda_1 \lambda_2 \lambda_3} \epsilon^{\lambda_1 \lambda_2 \lambda_3}{}_{\bar{\alpha}} \big)
\nonumber \\
+q \big({1 \over 8} H_{{\bar{\alpha}} \mu}{}^\mu -{i \over 3}  Y_{\lambda_1 \lambda_2 \lambda_3} \epsilon^{\lambda_1 \lambda_2 \lambda_3}{}_{\bar{\alpha}}  -{1 \over 8} \Phi_{\bar{\alpha}} \big) =0~,
\eea
\bea
\label{lin5}
p \big({1 \over 2} \Omega_{{\bar{\alpha}}, {\bar{\mu}}_1 {\bar{\mu}}_2}
-{1 \over 8} H_{{\bar{\alpha}} \lambda_1 \lambda_2} \epsilon^{\lambda_1 \lambda_2}{}_{{\bar{\mu}}_1 {\bar{\mu}}_2}
+{1 \over 16} \epsilon^\lambda{}_{{\bar{\alpha}} {\bar{\mu}}_1 {\bar{\mu}}_2} H_{\lambda \mu}{}^\mu
+{1 \over 16}  \epsilon^\lambda{}_{{\bar{\alpha}} {\bar{\mu}}_1 {\bar{\mu}}_2} \Phi_\lambda \big)
\nonumber \\
+q \big(-{1 \over 4} \Omega_{{\bar{\alpha}}, \lambda_1 \lambda_2}\epsilon^{\lambda_1 \lambda_2}{}_{{\bar{\mu}}_1 {\bar{\mu}}_2}
+{i \over 2} Y_{{\bar{\alpha}} \lambda_1 \lambda_2} \epsilon^{\lambda_1 \lambda_2}{}_{{\bar{\mu}}_1 {\bar{\mu}}_2}
+{1 \over 8} H_{{\bar{\alpha}} {\bar{\mu}}_1 {\bar{\mu}}_2} -{i \over 2} \epsilon^\lambda{}_{{\bar{\alpha}}
{\bar{\mu}}_1 {\bar{\mu}}_2} Y_{\lambda \mu}{}^\mu \big) =0~,
\eea
\bea
\label{lin6}
\partial_{\bar{\alpha}} q + q \big(-{i \over 2} \Lambda_{\bar{\alpha}} -{1 \over 4} h_{{\bar{\alpha}}}
-{1 \over 2} \Omega_{{\bar{\alpha}}, \mu}{}^\mu +i Y_{{\bar{\alpha}} \mu}{}^\mu \big)
+p \big(-{1 \over 4} H_{{\bar{\alpha}} \mu}{}^\mu -{1 \over 4} \Phi_{\bar{\alpha}} \big) =0~.
\eea

Similarly, the linear system associated with (\ref{alg3}) is
\bea
\label{lin7}
\frac{1}{12} p \epsilon_{\alpha}{}^{\bar\gamma_1\bar\gamma_2\bar\gamma_3}H_{\bar\gamma_1\bar\gamma_2\bar\gamma_3} -\frac{1}{4}q H_{\alpha\gamma}{}^\gamma -\frac{1}{4} q \Phi_\alpha + p \xi_\alpha = 0~,
\eea
\bea
\label{lin8}
\frac{1}{12}q \epsilon_{\bar\alpha}{}^{\gamma_1\gamma_2\gamma_3}H_{\gamma_1\gamma_2\gamma_3} + \frac{1}{4} p H_{\bar \alpha \gamma}{}^\gamma -\frac{1}{4}p  \Phi_{\bar\alpha} + q \xi_{\bar\alpha} = 0~.
\eea
The above system can be solved to express some of the fluxes in terms of the geometry, and find the conditions on the geometry imposed by supersymmetry.
In the analysis of the solutions, it is convenient to consider three different cases as described in section \ref{systsec}.

\subsection{Solution of the linear system}
\subsubsection{Generic Horizons}

The solution of the linear system can be arranged in different ways. The procedure which we adopt here is to solve first for the
complex field strengths and express them in terms of the real fields and functions $p,q$, and then use the remaining equations to find the
expression of the real fields in terms of the geometry, and to determine the conditions on the geometry. We shall demonstrate that, although
 the KSEs determine the complex fields in terms of the real fields and geometry, the real fields remain undetermined.

  To solve the linear system, first recall that in the generic case $p^2-q^2\not=0$ and $p^2+q^2=1$, and let us assume that in some open set $p,q\not=0$. If either $p,q$ vanish in an open set
   then the linear system will be solved as a special case. Next  we take the trace of (\ref{lin2}), and after a re-arrangement, that of (\ref{lin5}) to find
 \bea
 q\big(-\Omega_{\bar\l,}{}^{\bar\l}{}_{\a}+ i Y_{\a\l}{}^\l-{1\over8} \e_{\a}{}^{\bar\l_1\bar\l_2\bar\l_3} H_{\bar\l_1\bar\l_2\bar\l_3}\big)
 + p\big(-{1\over2} \e_{\a}{}^{\bar\l_1\bar\l_2\bar\l_3} \Omega_{\bar\l_1,\bar\l_2\bar\l_3}+{1\over8} H_{\a\l}{}^\l-{3\over8} \Phi_{\a}\big)=0~,
 \nonumber \\
 \la{trac1}
  \eea

  \bea
 p\big(\Omega_{\l,}{}^\l{}_{\bar\a}+ i Y_{\bar\a\l}{}^\l+{1\over8} \e_{\bar\a}{}^{\l_1\l_2\l_3} H_{\l_1\l_2\l_3}\big)
 + q \big({1\over2} \e_{\bar\a}{}^{\l_1\l_2\l_3} \Omega_{\l_1,\l_2\l_3}+{1\over8} H_{\bar\a\l}{}^\l+{3\over8} \Phi_{\bar\a}\big)=0~,
 \nonumber \\
 \la{trac2}
  \eea
 respectively.

  Next consider the equations (\ref{lin3}) and (\ref{lin7}), and (\ref{lin4}) and (\ref{lin8}) and solve them to express the components of the 3-form field strength $H$
  in terms of the rest of the fields yielding
  \bea
  {1\over8} H_{\a\l}{}^\l=-{1\over8} \Phi_{\a} +{pq\over2}  \xi_{\a} +p \partial_{\a} q-pq [{i\over2} \Lambda_{\a} +{1\over4} h_{\a}+{1\over2} \Omega_{\a,\l}{}^\l]
  +{i\over3} p^2  \e_{\a}{}^{\bar\l_1\bar\l_2\bar\l_3}Y_{\bar\l_1\bar\l_2\bar\l_3} ~,
  \cr
  {1\over24}\e_{\a}{}^{\bar\l_1\bar\l_2\bar\l_3} H_{\bar\l_1\bar\l_2\bar\l_3}=-{p^2\over2} \xi_{\a}+q\partial_{\a} q- q^2[{i\over2} \Lambda_{\a} +{1\over4} h_{\a}+{1\over2} \Omega_{\a,\l}{}^\l]
  +{i\over3} q p  \e_{\a}{}^{\bar\l_1\bar\l_2\bar\l_3}Y_{\bar\l_1\bar\l_2\bar\l_3} ~,
  \nonumber \\
  \la{hcompa}
  \eea
  \bea
  {1\over8} H_{\bar\a\l}{}^\l={1\over8} \Phi_{\bar\a} -{pq\over2}  \xi_{\bar\a} -q\partial_{\bar\a} p-pq [-{i\over2} \Lambda_{\bar\a} -{1\over4} h_{\bar\a}+{1\over2} \Omega_{\bar\a,\l}{}^\l]
  -{i\over3} q^2  \e_{\bar\a}{}^{\l_1\l_2\l_3}Y_{\l_1\l_2\l_3} ~,
  \cr
  {1\over24}\e_{\bar\a}{}^{\l_1\l_2\l_3}H_{\l_1\l_2\l_3}=-{q^2\over2} \xi_{\bar\a}+p\partial_{\bar\a} p+ p^2[-{i\over2} \Lambda_{\bar\a} -{1\over4} h_{\bar\a}+{1\over2} \Omega_{\bar\a,\l}{}^\l]
  +{i\over3} q p  \e_{\bar\a}{}^{\l_1\l_2\l_3}Y_{\l_1\l_2\l_3}~.
  \nonumber \\
  \la{hcomp}
  \eea
  It remains to solve the trace conditions (\ref{trac1}) and (\ref{trac2}) and (\ref{lin1}) and (\ref{lin6}) of the linear system in terms of $\phi$ and $\xi$.
  After some straightforward computation, one finds that
  \bea
 pq  \Phi_\a&=& \partial_\a q^2-(2+q^2) (i\Lambda_\a+{1\over2} h_\a)-(-2+5q^2) \Omega_{\a,\l}{}^\l-2 i (2-q^2) Y_{\a\l}{}^\l+ 2 q^2 \Omega_{\bar\l,}{}^{\bar\l}{}_\a
 \cr
 &&+2i pq \e_{\a}{}^{\bar\l_1\bar\l_2\bar\l_3} Y_{\bar\l_1\bar\l_2\bar\l_3}
 +pq \e_{\a}{}^{\bar\l_1\bar\l_2\bar\l_3} \Omega_{\bar\l_1,\bar\l_2\bar\l_3} ~,
\cr
p^2 q^2 \xi_\a&=&q^2 \partial_\a q^2- (1+2 q^4) [{i\over2} \Lambda_\a+{1\over4} h_\a]-{1\over2} (2 q^4+2 q^2-1) \Omega_{\a,\l}{}^\l-i Y_{\a\l}{}^\l+q^2 \Omega_{\bar\l,}{}^{\bar\l}{}_\a
\cr
&&+ pq (1+2q^2) {i\over3} \e_{\a}{}^{\bar\l_1\bar\l_2\bar\l_3} Y_{\bar\l_1\bar\l_2\bar\l_3}+{pq\over2}\e_{\a}{}^{\bar\l_1\bar\l_2\bar\l_3} \Omega_{\bar\l_1,\bar\l_2\bar\l_3} ~,
\cr
 pq \Phi_{\bar\a}&=&\partial_{\bar\a} p^2- (2+p^2) [i\Lambda_{\bar\a}+{1\over2} h_{\bar\a}]+ (-2+5 p^2) \Omega_{\bar\a,\l}{}^\l+ 2 i (2-p^2) Y_{\bar\a\l}{}^\l
   +2 p^2 \Omega_{\l,}{}^\l{}_{\bar\a}
   \cr
   &&+2i pq \e_{\bar\a}{}^{\l_1\l_2\l_3} Y_{\l_1\l_2\l_3} + pq  \e_{\bar\a}{}^{\l_1\l_2\l_3} \Omega_{\l_1, \l_2\l_3} ~,
   \cr
   p^2 q^2 \xi_{\bar\a}&=& p^2 \partial_{\bar\a} p^2-  (1+2p^4)[{i\over2}\Lambda_{\bar\a}+{1\over4} h_{\bar\a}]+{1\over2} (2 p^4+2p^2-1) \Omega_{\bar\a,\l}{}^\l+i Y_{\bar\a\l}{}^\l+ p^2 \Omega_{\l,}{}^\l{}_{\bar\a}
   \cr
   &&+ pq (1+2p^2) {i\over3} \e_{\bar\a}{}^{\l_1\l_2\l_3} Y_{\l_1\l_2\l_3}+{pq\over2}  \e_{\bar\a}{}^{\l_1\l_2\l_3} \Omega_{\l_1, \l_2\l_3}~.
   \eea
Clearly, we can substitute the $\Phi$'s and the $\xi$'s into (\ref{hcompa}) and (\ref{hcomp}) to express the traces and (3,0) and (0,3) part of $H$ in terms of the the $Y$ fluxes
and the geometry. However, we shall not do this here as for our conclusions this is not utilized.

Next using the expressions in (\ref{hcompa}) and (\ref{hcomp}), one can determine the (2,1) and (1,2) components of $H$ as
\bea
{p\over4} H_{\bar\a\mu_1\mu_2}&=& q[-{1\over2} \Omega_{\bar\a,\mu_1\mu_2}+i Y_{\bar\a\mu_1\mu_2}+i \delta_{\bar\a[\mu_1} Y_{\mu_2]\l}{}^\l]+{p\over4} \Omega_{\bar\a, \bar\nu_1\bar\nu_2} \e^{\bar\nu_1\bar\nu_2}{}_{\mu_1\mu_2}
\cr
&&-\delta_{\bar\a[\mu_1} \big[\partial_{\mu_2]}q- q \big({i\over2} \Lambda_{\mu_2]} +{1\over4} h_{\mu_2]}+{1\over2} \Omega_{\mu_2],\l}{}^\l\big)+{i\over3} p\, \e_{\mu_2]}{}^{\bar\l_1\bar\l_2\bar\l_3} Y_{\bar\l_1\bar\l_2\bar\l_3} \big] ~,
\cr
{q\over4} H_{\a\bar\mu_1\bar\mu_2}&=&-p\big[{1\over2} \Omega_{\a,\bar\mu_1\bar\mu_2}-i Y_{\a\bar\mu_1\bar\mu_2}+i \delta_{\a[\bar\mu_1} Y_{\bar\mu_2]\l}{}^\l\big]+{q\over4}
\Omega_{\a, \nu_1\nu_2} \e^{\nu_1\nu_2}{}_{\bar\mu_1\bar\mu_2}
\cr
&&-\delta_{\a[\bar\mu_1} \big[\partial_{\bar\mu_2]} p+p \big( -{i\over2} \Lambda_{\bar\mu_2]}-{1\over4} h_{\bar\mu_2]}+{1\over2} \Omega_{\bar\mu_2], \l}{}^\l\big)+{i q\over3} \e_{\bar\mu_2]}{}^{\l_1\l_2\l_3} Y_{\l_1\l_2\l_3}\big]~.
\nonumber \\
\eea
Note  that the expression  for $h$ in (\ref{hexp}) is not independent, and so if one substitutes the solution for the fluxes given above, one gets identically zero.
Conversely, any one of the solutions above can be used to express $h$ in terms of the fluxes and geometry.

Furthermore, one can use the solution of the linear system to determine $\mu$ in (\ref{muexp}) in terms of the $Y$ fluxes and geometry as
\bea
\sqrt 2\, pq \mu_{\bar\a}&=&-{i\over2} \Lambda_{\bar\a}+{1\over4} (p^2-q^2) h_{\bar\a}+{1\over2} (p^2-q^2) \Omega_{\bar\a,\l}{}^\l+i Y_{\bar\a\l}{}^\l
\cr
&&+ {2i\over 3}pq \e_{\bar\a}{}^{\l_1\l_2\l_3}
Y_{\l_1\l_2\l_3} ~.
\eea
Observe that the spin connection $\tilde \Omega$ of the horizon sections ${\cal S}$, which is given by $\tilde \Omega_{i,}{}^j{}_k=\Omega_{i,}{}^j{}_k$,  is not restricted. This indicates that there are no additional restrictions on the topology and geometry of ${\cal S}$ apart from those required for the global existence of certain forms which we describe in section \ref{systsec1}. This concludes the solution of the KSEs for the generic case.

The solution of the linear system for $Spin(7)$ horizons can be derived from that of the generic case after setting $p=q=1/\sqrt{2}$. This is straightforward to implement
and we shall not carry out the substitution here.

\subsubsection{Pure $SU(4)$ horizons}

For pure $SU(4)$ horizons, one has either $p=1, q=0$ or $p=0, q=1$.  The two cases are symmetric and without loss of generality, we choose $p=1, q=0$. A direct computation reveals that the solution  can be written as
\bea
\Phi_{\bar\a}= H_{\bar\a\l}{}^\l=0~,~~~
{1\over12} \e_\a{}^{\bar\l_1\bar\l_2\bar\l_3} H_{\bar\l_1\bar\l_2\bar\l_3}+\xi_\a=0~,
\cr
\Phi_\a=-\e_\a{}^{\bar\l_1\bar\l_2\bar\l_3} \Omega_{\bar\l_1,\bar\l_2\bar\l_3}+{2i\over3}\e_\a{}^{\bar\l_1\bar\l_2\bar\l_3} Y_{\bar\l_1\bar\l_2\bar\l_3}~,
\cr
H_{\a\l}{}^\l=e_\a{}^{\bar\l_1\bar\l_2\bar\l_3} \Omega_{\bar\l_1,\bar\l_2\bar\l_3}+ 2i \e_\a{}^{\bar\l_1\bar\l_2\bar\l_3} Y_{\bar\l_1\bar\l_2\bar\l_3}
\cr
{1\over8} \e_{\bar\a}{}^{\l_1\l_2\l_3} H_{\l_1\l_2\l_3} =-\Omega_{\l,}{}^\l{}_{\bar\a}-i Y_{\bar\a\l}{}^\l~,
\cr
{1\over2} H_{\bar\a\l_1\l_2}={1\over2} \Omega_{\bar\a,\bar\nu_1\bar\nu_2} \e^{\bar\nu_1\bar\nu_2}{}_{\l_1\l_2}-{2i\over3}
\delta_{\bar\a[\l_1} \e_{\l_2]}{}^{\bar\nu_1\bar\nu_2\bar\nu_3} Y_{\bar\nu_1\bar\nu_2\bar\nu_3}~,
\cr
-i Y_{\a\bar\l_1\bar\l_2}+{2i\over3} \delta_{\a[\bar\l_1} Y_{\bar\l_2]\mu}{}^\mu+{1\over2} \Omega_{\a,\bar\l_1\bar\l_2}
-{1\over3} \delta_{\a[\bar\l_1} \Omega_{|\mu|,}{}^\mu{}_{\bar\l_2]}=0~,
\cr
-{1\over2} h_{\bar\a}+{1\over3} \Omega_{\l,}{}^\l{}_{\bar\a}-{2i\over3} Y_{\bar\a\l}{}^\l=0~,
\cr
i \Lambda_{\bar\a}+ h_{\bar\a}- \Omega_{\bar\a,\l}{}^\l-\Omega_{\l,}{}^\l{}_{\bar\a}=0~. \label{puresol}
\eea

Moreover, one finds that
\bea\label{puremu}
\sqrt{2}\, \mu_{\bar\a}={1\over2} \e_{\bar\a}{}^{\l_1\l_2\l_3} \Omega_{\l_1,\l_2\l_3}+{i\over3}  \e_{\bar\a}{}^{\l_1\l_2\l_3} Y_{\l_1\l_2\l_3}~.
\eea
Again notice that there is no condition which involves only the spin connection $\tilde\Omega$ of the horizon section ${\cal S}$.  This indicates that
there are no additional conditions on the topology and geometry of the horizon sections apart from those required for the global existence of certain
forms which we describe in section \ref{systsec2}.

\subsection{Spinor conventions}

For our spinor conventions, we use those of \cite{sgiib}. In addition to integrate the KSEs along the lightcone directions, we have decomposed the
$Spin(9,1)$  spinors $\psi$ into positive and negative parts as
\be
\psi=\psi_+ + \psi_-, \qquad \Gamma_{\pm} \psi_{\pm}=0~,
\ee
which is a lightcone, or equivalently $Spin(1,1)$, chiral decomposition.  Furthermore, we have found the following identities useful
\bea
\Gamma_{\ell_1 \dots \ell_n} \psi_\pm &=& \pm (-1)^{[{n\over2}]} {1 \over (8-n)!} \epsilon_{\ell_1 \dots \ell_n}{}^{j_1 \dots j_{8-n}}
\Gamma_{j_1 \dots j_{8-n}} \psi_\pm~,~~~n\geq4~,
\eea
where $\psi$ is a positive chirality $Spin(9,1)$ spinor.  While, one has
\bea
\Gamma_{\ell_1 \dots \ell_n} \psi_\pm &=& \mp (-1)^{[{n\over2}]} {1 \over (8-n)!} \epsilon_{\ell_1 \dots \ell_n}{}^{j_1 \dots j_{8-n}}
\Gamma_{j_1 \dots j_{8-n}} \psi_\pm~,~~~n\geq4~,
\eea
provided that $\psi$ is a negative chirality $Spin(9,1)$ spinor.


\end{document}